\documentclass[a4paper,twocolumn,11pt]{quantumarticle}
\pdfoutput=1

\usepackage{enumitem}
\usepackage{graphicx}
\usepackage{array,booktabs,tabularx}
\usepackage{amsmath,amssymb,amsfonts}
\usepackage{tikz}
\usepackage{quantikz}
\usetikzlibrary{arrows.meta,positioning,calc}
\usepackage{xcolor}
\usepackage[utf8]{inputenc}
\usepackage[T1]{fontenc}
\usepackage{pgfplots}
\usepackage{subcaption}
\usepackage{amsthm}
\usepackage{url}
\usepackage{svg}

\usepackage{hyperref}
\usepackage{booktabs,graphicx}
\usepackage{tabularx}
\usepackage{listings} 
\usepackage{float}
\usepackage{ulem}
\hypersetup{
    colorlinks=true,
    linkcolor=blue,
    filecolor=magenta,      
    urlcolor=blue,
    citecolor=blue,
}

\pgfplotsset{compat=1.18}

\newcolumntype{Y}{>{\raggedright\arraybackslash}X}
\graphicspath{{figures/}}
\theoremstyle{definition}

\theoremstyle{plain}

\providecommand{\QFT}{\mathrm{QFT}}

\begin{document}

\title{A Quantum Bluestein's Algorithm for Arbitrary-Size Quantum Fourier Transform}
\author{Nan-Hong Kuo}
\email{d91222008@ntu.edu.tw}
\affiliation{Department of Physics, National Taiwan University, Taipei, Taiwan}

\author{Renata Wong}
\thanks{corresponding author.}
\email{renata.wong@cgu.edu.tw}
\affiliation{Department of Artificial Intelligence, Chang Gung University, Taoyuan, Taiwan}
\affiliation{Department of Neurology, Chang Gung Memorial Hospital, Keelung, Taiwan}
\maketitle

\begin{abstract}
We propose a quantum analogue of Bluestein's algorithm (QBA) that implements an exact $N$-point Quantum Fourier Transform (QFT) for arbitrary $N$. Our construction factors the $N$-dimensional QFT unitary $U_{\mathrm{QFT}_N}$ into three diagonal quadratic-phase gates and two standard radix-2 QFT subcircuits of size $M=2^m$ (with $M \ge 2N-1$). This achieves asymptotic gate complexity $O((\log N)^2)$ and uses $O(\log N)$ qubits, matching the performance of a power-of-two QFT on $m$ qubits while avoiding the need to embed into a larger Hilbert space. We validate the correctness of the algorithm through a concrete implementation in Qiskit and classical simulation, confirming that QBA produces the exact $N$-point discrete Fourier transform on arbitrary-length inputs.
\end{abstract}


\section{Introduction}

The Quantum Fourier Transform (QFT) is a core primitive in quantum computation, foundational to seminal algorithms in factoring and phase estimation~\cite{Shor1994,Nielsen2010}. While the canonical $n$-qubit circuit for the QFT is highly efficient, its applicability is strictly limited to transform sizes of $N=2^n$~\cite{Camps2021}. This constraint poses a significant challenge, as many practical applications require an exact transform for arbitrary integer sizes. The standard workaround of zero-padding an $N$-dimensional state into a larger power-of-two space is often inefficient and, more importantly, does not compute the true $N$-point discrete Fourier transform, but rather a finely sampled version of its spectrum~\cite{OppenheimSchafer2010}. This limitation is particularly acute in quantum machine learning applications, such as constructing kernel functions for SVMs, where exact propagator mapping requires flexible dimensionality \cite{KuoSVM}.

To address this limitation, we turn to a powerful technique from classical signal processing. Bluestein's algorithm, also known as the Chirp-$Z$ Transform, reformulates an arbitrary-size DFT as a convolution, which can then be solved efficiently using radix-2 Fast Fourier Transforms (FFTs)~\cite{Bluestein1970}. Here we introduce a direct quantum analogue: the Quantum Bluestein's Algorithm (QBA). Our construction provides a method to implement the exact $N$-point QFT for any integer $N$. The algorithm deterministically factors the target unitary $U_{\QFT_N}$ into a sequence of three diagonal quadratic-phase operators and two standard, highly-optimized radix-2 QFT subcircuits. This design retains the scaling of the standard QFT, achieving a gate complexity of $O(\log^2 N)$ on $O(\log N)$ qubits, while being universally applicable.

\paragraph{Related work}

The challenge of implementing the QFT for non-power-of-two sizes has been approached from several directions. Existing methods can be broadly classified into three families, each with distinct trade-offs.

The first family relies on mixed-radix and prime-factor factorizations, analogous to the classical Cooley-Tukey FFT~\cite{Cooley1965}. For a composite number $N$, the $\QFT_N$ can be decomposed into smaller QFTs corresponding to the prime factors of $N$. Circuit layouts based on the Chinese Remainder Theorem extend this idea, mapping the problem across co-prime sub-registers~\cite{Camps2021,Lei2024}. While these methods are exact and can be efficient for highly composite $N$, their circuit structure is irregular and depends heavily on the specific factorization of $N$, complicating uniform optimization and resource estimation.

A second approach is the Approximate Quantum Fourier Transform (AQFT). The AQFT systematically truncates the controlled-phase rotations in the standard QFT circuit, omitting those with angles smaller than a predefined threshold~\cite{Coppersmith2002AQFT}. This significantly reduces the gate count and circuit depth, especially the number of two-qubit gates, at the cost of introducing a controllable error~\cite{Nam2020}. While powerful for applications where precision can be traded for resources, it is fundamentally an approximation.

The third, and most direct, approach is to simply embed the $N$-dimensional problem into a larger, $M=2^m$ dimensional space and apply the standard $\QFT_M$. This zero-padding technique is simple to implement but, as established in classical signal processing, it does not compute the true $N$-point DFT. Instead, it computes more samples of the signal's Z-transform on the unit circle, which is a different mathematical operation~\cite{OppenheimSchafer2010}.

Our Quantum Bluestein's Algorithm offers a complementary and distinct path. It is exact for any integer $N$, prime or composite. Rather than factoring the problem based on the arithmetic properties of $N$, it reduces the $\QFT_N$ to a convolution that is then solved using structurally uniform, highly-optimized radix-2 QFT subroutines. This makes QBA a general-purpose tool that offers predictable resource scaling and reuses some of the most well-understood components in quantum computing.

\section{Quantum Bluestein's Algorithm \label{sec:qba}}

Our quantum construction is a direct analogue of the classical Bluestein's algorithm, which provides an efficient method for computing the $N$-point DFT for any integer $N$. The classical algorithm begins with the standard DFT definition
$$y_k = \sum_{j=0}^{N-1} x_j e^{-2\pi i j k / N}$$
and employs the algebraic identity 
$$2jk = j^2 + k^2 - (k-j)^2$$
Substituting this into the exponent transforms the DFT sum into a convolution:
\begin{align*}
    y_k &= \sum_{j=0}^{N-1} x_j \exp\left( \frac{\pi i}{N} \left[ -(k-j)^2 + j^2 + k^2 \right] \right) \\
    &= \underbrace{e^{\frac{\pi i}{N} k^2}}_{\text{output de-chirp}} \sum_{j=0}^{N-1} \left( \underbrace{x_j e^{\frac{\pi i}{N} j^2}}_{\text{input chirp } a_j} \right) \underbrace{e^{-\frac{\pi i}{N} (k-j)^2}}_{\text{kernel } b_{k-j}}
    \label{eq:bluestein_conv}
\end{align*}

This expression decomposes the DFT into three steps: (i) multiplying the input signal by a quadratic phase factor (a "chirp"), (ii) convolving the result with a chirp kernel, and (iii) multiplying the output by a final de-chirp factor. By the convolution theorem, this process can be efficiently implemented using power-of-two FFTs of size $M \ge 2N - 1$.

\subsection{Quantum circuit construction}

We now construct QBA, a quantum algorithm that implements this logic.
The QBA operates on an $m$-qubit register, where $m$ is chosen such that $M = 2^m \ge 2N - 1$. This ensures a workspace of dimension $M$ (a power-of-two) large enough to perform the convolution without aliasing, analogous to the zero-padding step in the classical algorithm. The input is a quantum state 
\[ 
|\psi_{\text{in}}\rangle = \sum_{j=0}^{N-1} x_j\,|j\rangle
\] 
assuming that any basis states $|j\rangle$ with $j \ge N$ start with zero amplitude. 
The algorithm proceeds in five main steps, as illustrated in Fig.~\ref{fig:qba_circuit} and laid out below: 

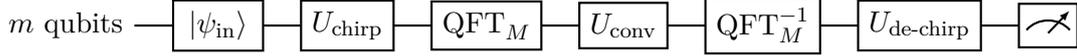
\begin{figure*}[t]
\centering
\begin{quantikz}[row sep=0.2cm]
\lstick{$m$ qubits} & \gate{\,|\psi_{\text{in}}\rangle\,} & \gate{U_{\text{chirp}}} & \gate{\text{QFT}_M} & \gate{U_{\text{conv}}} & \gate{\text{QFT}_M^{-1}} & \gate{U_{\text{de-chirp}}} & \meter{}
\end{quantikz}
\caption{High-level circuit diagram for the Quantum Bluestein's Algorithm (QBA). The algorithm computes the arbitrary-size QFT by sequentially applying an input chirp, a standard power-of-two QFT, a Fourier-domain convolution operator, an inverse QFT, and an output de-chirp.}
\label{fig:qba_circuit}
\end{figure*}


\paragraph{1. Input chirp} Apply a diagonal unitary operator $U_{\text{chirp}}$ to the input state. This multiplies the amplitude of each basis state $|j\rangle$ by a quadratic phase (a “chirp” factor), where $j \in \{0,\ldots,M-1\}$:
$$ U_{\text{chirp}}\,|j\rangle = e^{-\pi i j^2/N}\,|j\rangle$$
This step prepares the quantum analogue of the sequence $a_j$ from Bluestein’s classical algorithm. The state becomes:
\begin{equation*}
\ket{\psi_1} = U_{\text{chirp}}\ket{\psi_{\text{in}}} = \sum_{j=0}^{M-1} a_j \ket{j}
\end{equation*} 
where $a_j = x_je^{\frac{\pi i}{N}j^2}$.

\paragraph{2. Forward QFT} Apply a standard $m$-qubit QFT of length $M=2^m$ to the chirped state. This transforms $|\psi_1\rangle$ into the Fourier basis. Let $|\psi_{2}\rangle$ denote the state of the system in the Fourier basis:
\begin{equation*}
    |\psi_{2}\rangle = \sum_{k=0}^{M-1} \tilde{a}_k |k\rangle
\end{equation*}
where $\tilde{a}_k$ represents the Fourier coefficients of the chirped input sequence $a_j$.

\paragraph{3. Convolution in the Fourier basis}

The objective of this step is to multiply the coefficients $\tilde{a}_k$ element-wise by the Fourier transform of the chirp kernel. Let $b_j = e^{\pi i j^2 / N}$ be the chirp kernel sequence, and let $\tilde{b}_k$ denote the $k$-th component of its $M$-point DFT, i.e., $\tilde{b} = \text{DFT}_M(b)$.

We aim to apply a diagonal operator $U_{conv}$ such that:
\begin{equation*}
    U_{conv} |k\rangle = \tilde{b}_k |k\rangle, \quad k \in \{0, \dots, M-1\}
\end{equation*}

Unlike the input chirp $U_{chirp}$, the values $\tilde{b}_k$ are complex numbers that generally do not have unit magnitude ($|\tilde{b}_k| \neq 1$). Consequently, the diagonal operator $U_{conv}$ is not unitary. To implement this operation, we embed it into a larger unitary $\tilde{U}_{conv}$ by introducing a single ancilla qubit initialized to $|0\rangle$.

Using a block-encoding approach, we construct $\tilde{U}_{conv}$ to act on the joint state $|k\rangle \otimes |0\rangle$ as follows:
\begin{equation*}
    \tilde{U}_{conv}: |k\rangle \otimes |0\rangle \mapsto |k\rangle \otimes \left( \frac{\tilde{b}_k}{\alpha} |0\rangle + \sqrt{1 - \left|\frac{\tilde{b}_k}{\alpha}\right|^2} |1\rangle \right)
\end{equation*}
Here, $\alpha$ is a suitable normalization constant (where $\alpha \ge \max_k |\tilde{b}_k|$) ensuring the transformation is unitary.

Applying this to the superposition $|\psi_2\rangle \otimes |0\rangle$, the system evolves to:
\begin{equation*}
    |\psi_3\rangle = \sum_{k=0}^{M-1} \tilde{a}_k |k\rangle \otimes \left( \frac{\tilde{b}_k}{\alpha} |0\rangle + \ket{\text{garbage}}_k \right)
\end{equation*}
The successful branch of the computation corresponds to the subspace where the ancilla is in state $|0\rangle$. Within this subspace, the amplitudes of the main register are effectively multiplied by $\tilde{b}_k$ (scaled by $1/\alpha$), implementing the required convolution multiplication in the Fourier domain. In other words, we postselect on the ancilla being in state $|0\rangle$.

\paragraph{4. Inverse QFT}

Next, apply the inverse QFT ($QFT_M^{-1}$) to the main register. This transforms the state from the Fourier basis back to the computational basis. By the convolution theorem, this operation yields the linear convolution of the input sequence $a$ and the kernel $b$:
\begin{equation*}
    QFT_M^{-1} |\psi_{main}\rangle \propto \sum_{j=0}^{M-1} (a * b)_j |j\rangle
\end{equation*}
where $\ket{\psi_\text{main}}$ is the post-selected state of the main register. 

\paragraph{5. Output de-chirp}

Finally, we apply a diagonal unitary $U_{de-chirp}$ to the register. This multiplies each basis state $|k\rangle$ by a quadratic phase factor to cancel the chirp introduced in Step 1:
\begin{equation*}
    U_{de-chirp} |k\rangle = e^{-\pi i k^2 / N} |k\rangle, \quad k \in \{0, \dots, M-1\}
\end{equation*}

The resulting final state $|\psi_{out}\rangle$ encodes the exact $N$-point DFT of the original input coefficients $\{x_j\}$ in its amplitudes (for indices $k < N$). Measuring the register in the computational basis will therefore yield outcome $k$ with probability $|\langle k|\psi_{\text{out}}\rangle|^2$, corresponding to the $|y_k|^2$ of the desired Fourier spectrum.


\subsection{Complexity analysis}
The efficiency of the QBA depends on the efficient implementation of the three diagonal unitary operators ($U_{\mathrm{chirp}}$, $U_{\mathrm{conv}}$, $U_{\mathrm{de\text{-}chirp}}$) and the two standard QFTs.

The implementation of diagonal unitary operators with quadratic phases, such as $U_{\text{chirp}} = \text{diag}(e^{-i\pi j^2/N})$, relies on the efficient computation of the function $f(j) = j^2$ in the phase exponent. By expanding the integer index $j$ into its binary representation $j = \sum_{l=0}^{m-1} j_l 2^l$, the square term transforms into a sum of bitwise products:
\begin{equation*}
    j^2 = \sum_{l=0}^{m-1} j_l 2^{2l} + \sum_{0 \le l < r \le m-1} j_l j_r 2^{l+r+1}
\end{equation*}
This decomposition indicates that the global quadratic phase can be synthesized exactly using a sequence of single-qubit $R_Z$ rotations (for the $j_l$ self-terms, noting $j_l^2=j_l$ for bits) and two-qubit controlled-$R_Z$ rotations (for the $j_l j_r$ cross-terms). The total number of gates required corresponds to the number of terms in this expansion, yielding exactly $m(m+1)/2$ rotation gates, which scales as $\mathcal{O}(m^2)$ \cite{zalka}. Consequently, the phase oracle does not require full arithmetic computation of the square into an ancillary register, but rather can be realized directly via controlled rotations on the computational basis states.


The standard $m$-qubit QFT and its inverse are known to have a gate complexity of $\mathcal{O}(m^2)$. Since all five steps of the QBA can be implemented with $\mathcal{O}(m^2)$ gates, and $m = \lceil \log_2(2N-1) \rceil = \mathcal{O}(\log N)$, the total gate complexity of the QBA is $\mathcal{O}((\log N)^2)$. The qubit requirement is $m$ primary qubits plus a small number of ancillary qubits for arithmetic, totaling $\mathcal{O}(\log N)$. The complexity is summarized in Table~\ref{tab:qba_complexity}. 


\begin{table}[h]
    \centering
    \caption{Complexity analysis of QBA.}
    \label{tab:qba_complexity}
    \resizebox{\linewidth}{!}{
    \begin{tabular}{@{}lcc@{}}
    \toprule
    \textbf{Component} & \textbf{Gate complexity} & \textbf{Qubit requirement} \\
    \midrule
    Input/Output chirp operators & $\mathcal{O}((\log N)^2)$ & $\mathcal{O}(\log N)$ \\
    Convolution operator & $\mathcal{O}((\log N)^2)$ & $\mathcal{O}(\log N)$ \\
    QFT / inverse QFT (size $M$) & $\mathcal{O}((\log N)^2)$ & $\mathcal{O}(\log N)$ \\
    \midrule
    \textbf{Total QBA} & $\boldsymbol{\mathcal{O}((\log N)^2)}$ & $\boldsymbol{\mathcal{O}(\log N)}$ \\
    \bottomrule
    \end{tabular}}
\end{table}

This result confirms that QBA is asymptotically as efficient as the standard radix-2 QFT, while removing the critical restriction on the input size $N$.

\section{Proof of correctness}

Here we present a proof that the proposed algorithm is in fact correct and its output corresponds exactly to the expected output of DFT on the first $N$ qubit states. 

\textbf{Theorem 2.1.} \textit{Let $N \in \mathbb{Z}_{>0}$ be the target transform size and $M = 2^m \ge 2N-1$ be the computational workspace size. The quantum circuit defined by the operator sequence $U_{QBA} = U_{de\text{-}chirp} \cdot QFT_M^{-1} \cdot U_{conv} \cdot QFT_M \cdot U_{chirp}$ exactly implements the $N$-point Discrete Fourier Transform on the subspace spanned by the first $N$ computational basis states.}

\begin{proof}
Let the input state $|\psi_{in}\rangle$ be a superposition of the first $N$ basis states with coefficients $x_j$, embedded into the larger Hilbert space $\mathcal{H}_M$ (zero-padding):
\begin{equation*}
    |\psi_0\rangle = \sum_{j=0}^{M-1} x_j |j\rangle, \quad \text{where } x_j = 0 \text{ for } j \ge N.
\end{equation*}
The algorithm proceeds in three phases corresponding to the Bluestein decomposition:

\textbf{1. Input chirp}
Applying the diagonal unitary $U_{chirp} = \text{diag}(e^{-\frac{\pi i}{N}j^2})$ to $|\psi_0\rangle$ yields:
\begin{equation*}
    |\psi_1\rangle = U_{chirp}|\psi_0\rangle = \sum_{j=0}^{N-1} x_j e^{-\frac{\pi i}{N}j^2} |j\rangle.
\end{equation*}
Let $a_j = x_j e^{-\frac{\pi i}{N}j^2}$ denote the chirped input sequence.

\textbf{2. Linear convolution via unitary convolution}
The sequence $QFT_M^{-1} \cdot U_{conv} \cdot QFT_M$ implements the circular convolution of the input amplitudes with the kernel vector $b$, defined by $b_t = e^{\frac{\pi i}{N}t^2}$ for $t \in \{0, \dots, M-1\}$. By the Convolution Theorem:
\begin{equation*}
    |\psi_{conv}\rangle = \sum_{k=0}^{M-1} (a * b)_k |k\rangle,
\end{equation*}
where $(a * b)_k = \sum_{j=0}^{M-1} a_j b_{(k-j) \text{ mod } M}$.
Since the input sequence $a_j$ is non-zero only for $0 \le j < N$ and the workspace size satisfies $M \ge 2N-1$, the circular convolution is equivalent to the linear convolution for the first $N$ indices ($0 \le k < N$). Thus, for $k < N$:
\begin{equation*}
    (a * b)_k = \sum_{j=0}^{N-1} a_j b_{k-j} = \sum_{j=0}^{N-1} \left( x_j e^{-\frac{\pi i}{N}j^2} \right) e^{\frac{\pi i}{N}(k-j)^2}.
\end{equation*}

\textbf{3. Output de-chirp}
Finally, we apply $U_{de\text{-}chirp} = \text{diag}(e^{-\frac{\pi i}{N}k^2})$. The amplitude of the basis state $|k\rangle$ (for $k < N$) becomes:
\begin{equation*}
    y_k = e^{-\frac{\pi i}{N}k^2} \cdot (a * b)_k = e^{-\frac{\pi i}{N}k^2} \sum_{j=0}^{N-1} x_j e^{-\frac{\pi i}{N}j^2} e^{\frac{\pi i}{N}(k-j)^2}
\end{equation*}
Substituting the expansion $(k-j)^2 = k^2 - 2jk + j^2$ into the exponent:
\begin{align*}
    \text{Exponent} &= -\frac{\pi i}{N}k^2 - \frac{\pi i}{N}j^2 + \frac{\pi i}{N}(k^2 - 2jk + j^2) \\
    &= \frac{\pi i}{N} \left( -k^2 - j^2 + k^2 - 2jk + j^2 \right) \\
    &= -\frac{2\pi i}{N}jk.
\end{align*}
Substituting this back into the expression for $y_k$:
\begin{equation*}
    y_k = \sum_{j=0}^{N-1} x_j e^{-\frac{2\pi i}{N}jk}.
\end{equation*}
This is exactly the definition of the $N$-point Discrete Fourier Transform. Thus, for any $k \in \{0, \dots, N-1\}$, the amplitude of $|k\rangle$ in the final state corresponds exactly to the $k$-th Fourier component of the input vector $x$.
\end{proof}

\subsection{Simulation using Qiskit}
\label{sec:qiskit-simulation}

Beyond the algebraic proof that $U_{\mathrm{QFT}_N}$ factorizes into three quadratic-phase diagonals and two radix-2 QFT subroutines, we instantiate a concrete, gate-level realization in Qiskit. The circuit on $m=\lceil \log_2 M\rceil$ qubits (with $M\ge 2N-1$) stacks
\begin{equation*}
U_{\mathrm{de\text{-}chirp}}\;\cdot\;\mathrm{QFT}_M^{-1}\;\cdot\;U_{\mathrm{conv}}\;\cdot\;\mathrm{QFT}_M\;\cdot\;U_{\mathrm{chirp}},
\end{equation*}
where each quadratic-phase diagonal is compiled as a product of single- and two-qubit $R_Z$ / controlled-$R_Z$ rotations coming from the bit-quadratic expansion of $j^2$ (self and pairwise terms). This yields exactly $m(m+1)/2$ primitive $Z$-rotation angles per diagonal block. 



\section{Conclusion}
\label{sec:conclusion}
The proposed quantum version of Bluestein's algorithm resolves the power-of-two restriction of the QFT. The asymptotic complexity of the algorithm matches that of QFT, both in terms of gate count and number of qubits.

\section{Code availability \label{code}}
The source code is publicly available at  \url{https://github.com/renatawong/quantum-bluestein}. The user may choose to execute the quantum circuit using either a simulator or a real quantum backend. For comparison, the code implements also the classical Bluestein's algorithm. 

\appendix
\section{Worked examples}

\paragraph{Notation} 
Throughout the paper, $N$ denotes the target transform length of $\mathrm{QFT}_N$ (i.e., the DFT size).
To implement Bluestein’s reduction without aliasing, we embed into a power-of-two workspace of size
$M=2^m$ chosen such that $M\ge 2N-1$; here $m=\lceil\log_2 M\rceil$ is the number of physical qubits.
Thus, $N$ is a problem-size parameter independent of the hardware register size $m$;
for example, $N=3$ uses $m=3$ qubits ($M=8$), and $N=6$ uses $m=4$ qubits ($M=16$).

\subsection{Example: $N=3, M=8$}

To verify the algorithm, we apply the QFT to a basis state $\ket{j}$ with $N=3$. We demonstrate how this logical problem is embedded into a hardware register of 3 qubits.

We use a 3-qubit system ($m=3$), creating a Hilbert space of size $M=2^3=8$. The computational basis states are $\ket{000}, \ket{001}, \dots, \ket{111}$.

\paragraph{Target state}
Let's say we want to compute the QFT on the input state $\ket{j} = \ket{001}$. For $x_1=1$, the expected DFT output is $y_k = e^{-\frac{2\pi i}{3} (1) k} = \omega_3^k$.
\begin{align*}
    k=0 &: \quad y_0 = 1 \\
    k=1 &: \quad y_1 = e^{-i 2\pi / 3} \\
    k=2 &: \quad y_2 = e^{-i 4\pi / 3}
\end{align*}
The target state is defined only on the subspace spanned by the first $N$ basis states $\{\ket{0}, \ket{1}, \ket{2}\}$:
\begin{equation} \label{eq:target}
    \ket{\psi_{\text{t}}} = \frac{1}{\sqrt{3}} \left( \ket{0}_{\!L} + e^{-i \frac{2\pi}{3}} \ket{1}_{\!L} + e^{-i \frac{4\pi}{3}} \ket{2}_{\!L} \right)
\end{equation}
where $\ket{0}_{\!L}=\ket{000}$, $\ket{1}_{\!L}=\ket{001}$, and $\ket{2}_{\!L}=\ket{010}$.

\paragraph{Input vector} We verify the algorithm using the standard basis state $\ket{1}$. We initialize the 3-qubit register in the binary state representing integer 1.
\begin{equation}
    \ket{\psi_0} = \ket{001}
\end{equation}
To perform linear convolution of size $N=3$ using a cyclic convolution method, we pad the input to size $M \geq 2N-1 = 5$. The nearest power of 2 is $M=8$ (requiring 3 qubits). $\ket{\psi_0}$ naturally pads the input, as amplitudes for basis states $\ket{011}$ through $\ket{111}$ are zero. 

\paragraph{Input chirp gate}
We apply a diagonal phase gate $D_{in}$ where the phase depends on the integer index $j$. For state $\ket{001}$ ($j=1$), the phase is 
\[ e^{-\frac{\pi i}{3} j^2} = e^{-i \pi/3} \]
and hence 
\begin{equation}
    \ket{\psi_1} = e^{-i \pi/3} \ket{001}
\end{equation}


\paragraph{Unitary convolution}
The circuit performs a convolution of the amplitude vector with a kernel vector $b$ using QFTs of size $M=8$. The input amplitude at index 1 is $a_1 = e^{-i \pi / 3}$, while the kernel vector components are $b_{k-j} = e^{\frac{\pi i}{3} (k-j)^2}$.

The convolution $(a * b)_k$ determines the amplitude of the basis state $\ket{k}$:
\begin{align*}
    \ket{000} (k=0): & \quad a_1 \cdot b_{-1} = e^{-i \pi/3} \cdot e^{i \pi/3} = 1 \\
    \ket{001} (k=1): & \quad a_1 \cdot b_{0} = e^{-i \pi/3} \cdot 1 = e^{-i \pi/3} \\
    \ket{010} (k=2): & \quad a_1 \cdot b_{1} = e^{-i \pi/3} \cdot e^{i \pi/3} = 1
\end{align*}
The register evolves to a superposition of 3-qubit states:
\begin{equation*}
    \ket{\psi_2} = \alpha \left( \ket{000} + e^{-i \pi/3}\ket{001} + \ket{010} + \ket{\text{g}} \right)
\end{equation*}
where the $\ket{\text{g}}$ occupies states $\ket{011} \dots \ket{111}$ which are outside the logical $N=3$ subspace.

\begin{figure}[h]
\centering
\includegraphics[width=0.5\textwidth]{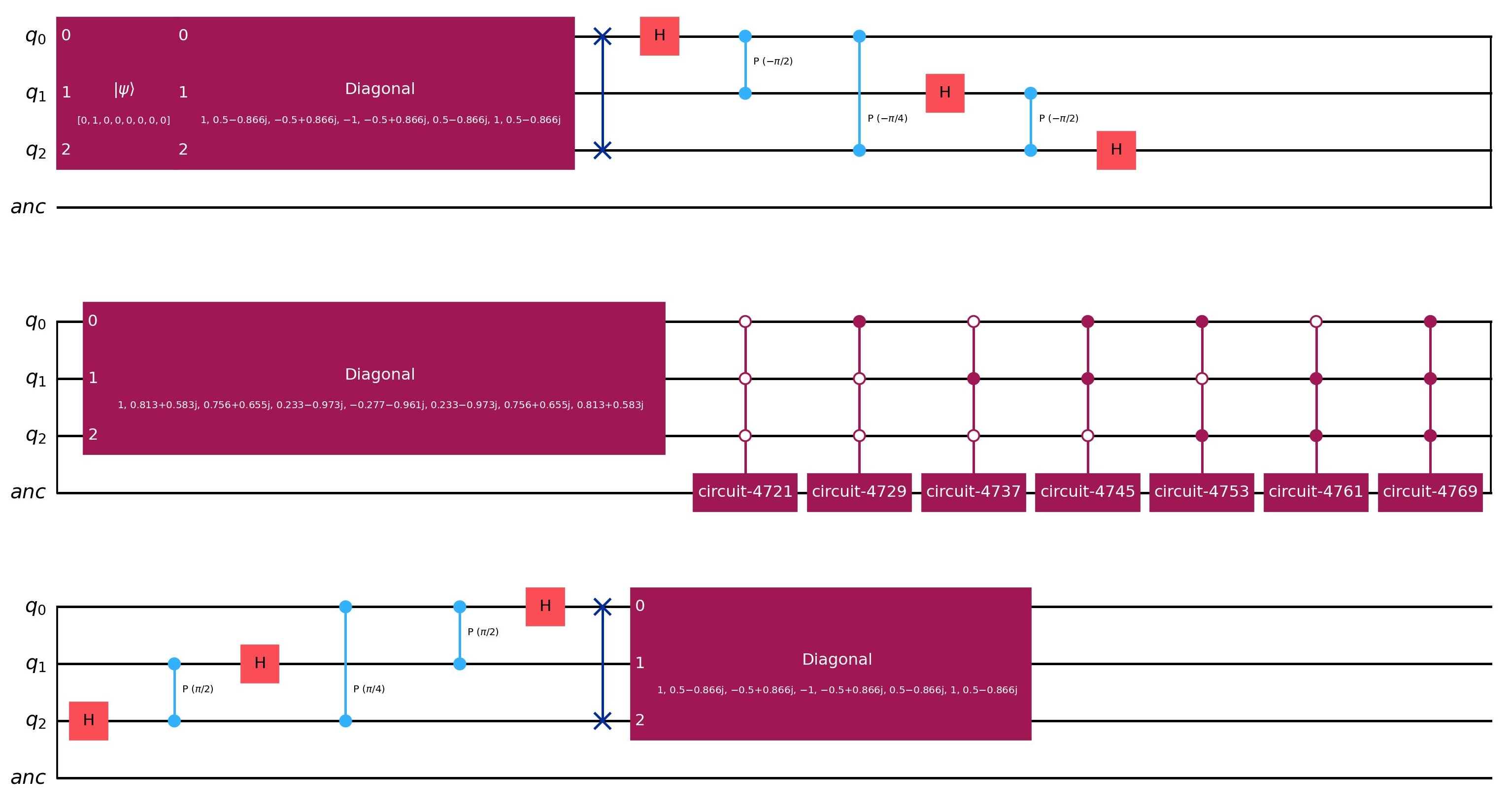}
\caption{Quantum circuit for $N=3$ generated using QBA.}
\label{fig-qba-n3}
\end{figure}

\begin{figure}[h]
\centering
\includegraphics[width=0.4\textwidth]{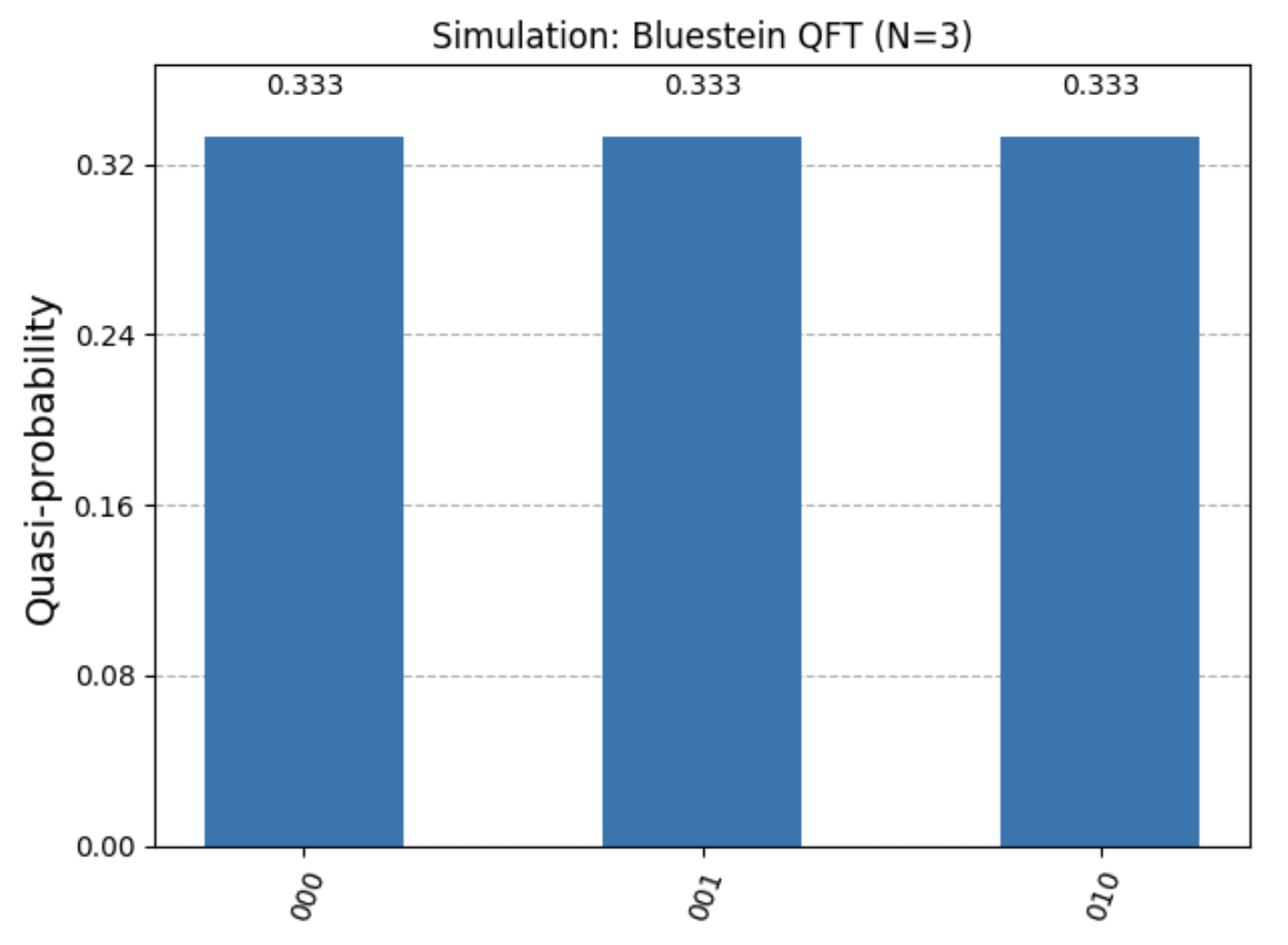}
\caption{Histogram of probabilities for $N=3$ generated using QBA.}
\label{fig-qba-n3-hist}
\end{figure}

\paragraph{Output de-chirp gate}
We apply the diagonal phase gate $D_{out}$ with phases $e^{-\frac{\pi i}{3} k^2}$ to the 3-qubit basis states:
\begin{itemize}
    \item $\ket{000} \xrightarrow{k=0} \text{phase } 1$.
    \item $\ket{001} \xrightarrow{k=1} \text{phase } e^{-i \pi/3}$.
    \item $\ket{010} \xrightarrow{k=2} \text{phase } e^{-i 4\pi/3}$.
\end{itemize}
The final state is:
\begin{align*}
    \ket{\psi_{f}} &= \alpha \left( \ket{000} + e^{-i \frac{2\pi}{3}} \ket{001} + e^{-i \frac{4\pi}{3}} \ket{010} + \ket{g} \right)
\end{align*}
Within the valid logical subspace (the first $N=3$ computational basis states), the output matches the target QFT state exactly. The output generated by our algorithm using Qiskit and the state vector simulator (see Section~\ref{code}) for this example is: 
\begin{equation*}
[0.57735 +0.j, -0.288675-0.5j, -0.288675+0.5j]
\end{equation*}
Execution on a real device leads unavoidably to a superposition state due to high level of noise present in real systems, and doesn't therefore correctly reflect the potential of the algorithm. Fig.~\ref{fig-qba-n3} shows the corresponding circuit, while Fig.~\ref{fig-qba-n3-hist} shows the histogram. 

\subsection{Example: $N=6, M=16$}
We illustrate the construction on a non-power-of-two, non-prime case.

To implement Bluestein's reduction without aliasing, we embed the problem into a power-of-two workspace $M$ such that $M \ge 2N - 1 = 11$. The nearest power of two is $M=16$ (requiring $m=4$ qubits). The computational basis states range from $|0000\rangle$ to $|1111\rangle$. The valid logical subspace corresponds to the first $N=6$ basis states $\{|0\rangle, \dots, |5\rangle\}$.

\paragraph{Input vector}
We choose the input vector $x$ such that the first three indices ($j=0, 1, 2$) have unit amplitude:
\begin{equation*}
    x = [1, 1, 1, 0, 0, 0]
\end{equation*}
The corresponding quantum state (unnormalized for clarity) is:
\begin{equation*}
    |\psi_{in}\rangle = |0\rangle + |1\rangle + |2\rangle = |0000\rangle + |0001\rangle + |0010\rangle
\end{equation*}

\paragraph{Target output (DFT)}
For the given input, the expected $N$-point DFT output $y_k$ is given by the geometric series 
$$y_k = \sum_{j=0}^{5} x_j \cdot e^{-\frac{2\pi i}{6}jk}$$ 
We calculate the first two terms:
\begin{align*}
    k=0 &: y_0 = 3 \\
    k=1 &: y_1 = 1 + e^{-i\pi/3} + e^{-2\pi i/3} = 1 - i\sqrt{3} \\
    k=2 &: y_2 = 0 \\
    k=3 &: y_3 = 1 \\
    k=4 &: y_4 = 0 \\
    k=5 &: y_5 = 1 + i\sqrt{3} \\
\end{align*}
\paragraph{Input chirp gate}
We apply the diagonal unitary $U_{chirp}$ where the phase is $e^{-\frac{\pi i}{N}j^2}$. The calculated phases are as follows for a particular $j$:
\begin{align*}
    j=0 &: 1 \\
    j=1 &: e^{-i\pi/6} \\
    j=2 &: e^{-2\pi i/3} \\
    j=3 &: i \\
    j=4 &: e^{-2\pi i/3} \\
    j=5 &: e^{-\pi i/6} \\
\end{align*}

The state transforms to the sequence $a_j$:
\begin{equation*}
    |\psi_{1}\rangle = \sum_{j=0}^5 e^{-\frac{\pi*i}{6}j^2}\ket{j}
\end{equation*}

\paragraph{Unitary convolution}
The circuit performs the convolution $(a * b)_k$ using QFTs of size $M=16$, where the kernel is $b_t = e^{\frac{\pi i}{6}t^2}$. We compute the convolution for the first two output indices ($k=0, 1$) to verify the mechanism:

For $k=0$, the convolution sum is $\sum_{j} a_j b_{0-j}$:
\begin{align*}
    j=0 &: a_0 b_0 = 1 \cdot 1 = 1 \\
    j=1 &: a_1 b_{-1} = e^{-i\pi/6} \cdot e^{\frac{\pi i}{6}(-1)^2} = e^{-i\pi/6} \cdot e^{i\pi/6} = 1 \\
    j=2 &: a_2 b_{-2} = e^{-2\pi i/3} \cdot e^{\frac{\pi i}{6}(-2)^2} = 1 \\
    \text{Sum} &: 1 + 1 + 1 = 3
\end{align*}

For $k=1$, the convolution sum is $\sum_{j} a_j b_{1-j}$:
\begin{align*}
    j=0 &: a_0 b_1 = 1 \cdot e^{\frac{\pi i}{6}(1)^2} = e^{i\pi/6} \\
    j=1 &: a_1 b_0 = e^{-i\pi/6} \cdot 1 = e^{-i\pi/6} \\
    j=2 &: a_2 b_{-1} = e^{-2\pi i/3} \cdot e^{\frac{\pi i}{6}(-1)^2} = -i \\
    \text{Sum} &: e^{i\pi/6} + e^{-i\pi/6} - i = \sqrt{3} - i
\end{align*}

\paragraph{Output de-chirp gate}
Finally, we apply the diagonal phase $U_{de-chirp}$ with phases $e^{-\frac{\pi i}{6}k^2}$.
\begin{itemize}
    \item For $k=0$, the result is $3 \cdot e^{-0} = 3$
    \item For $k=1$, the result is $1 - i\sqrt{3}$
\end{itemize}
Both match the respective targets $y_0$ and $y_1$. 

The final state is 
\begin{equation*}
\ket{\psi_f} = \frac{1}{\sqrt{18}} (3\ket{0}+(1-i\sqrt{3})\ket{1}+\ket{3}+(1+i\sqrt{3})\ket{5})
\end{equation*}

\begin{figure}[h]
\centering
\includegraphics[width=0.5\textwidth]{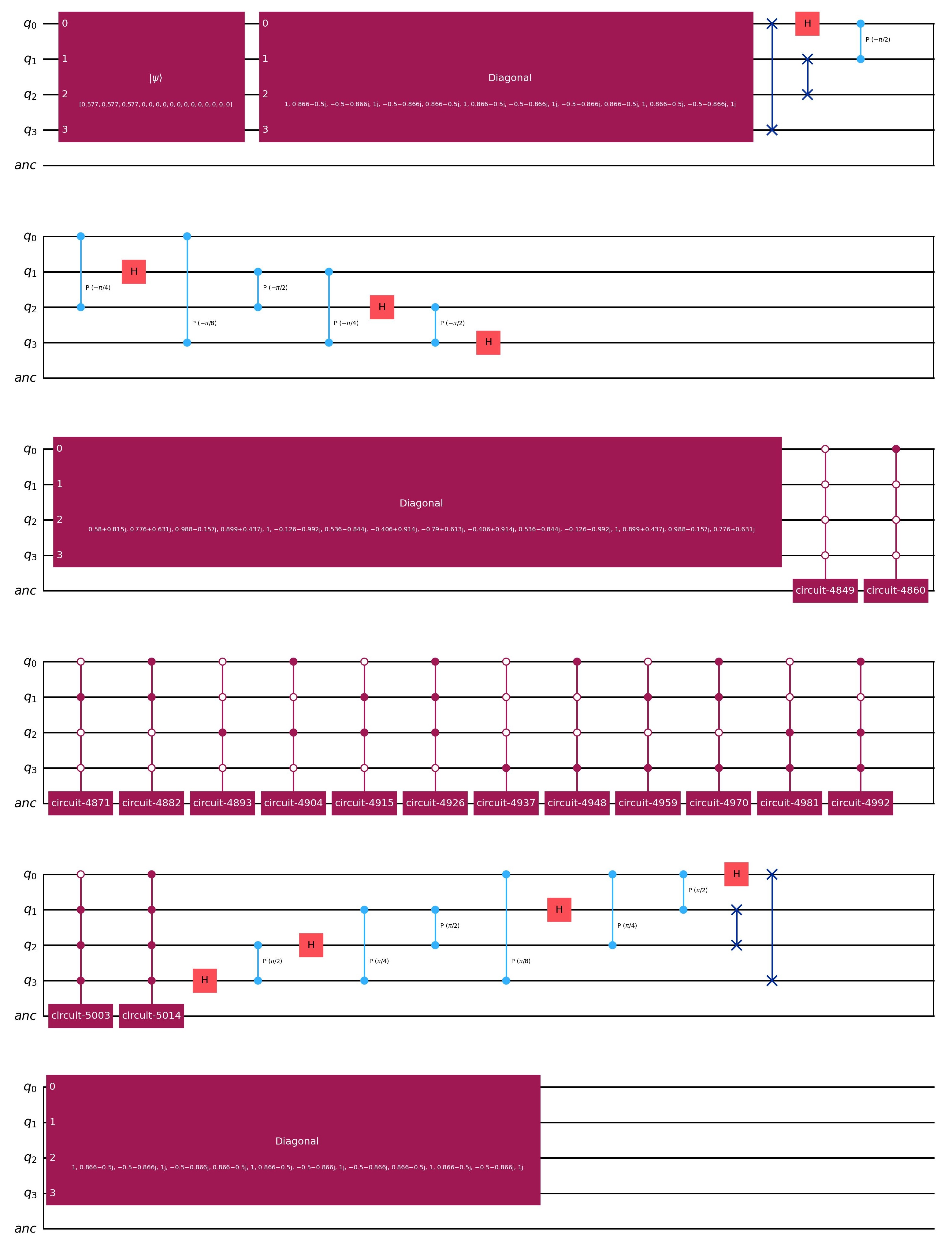}
\caption{Quantum circuit for $N=6$ generated using QBA.}
\label{fig-qba-n6}
\end{figure}

\begin{figure}[h]
\centering
\includegraphics[width=0.4\textwidth]{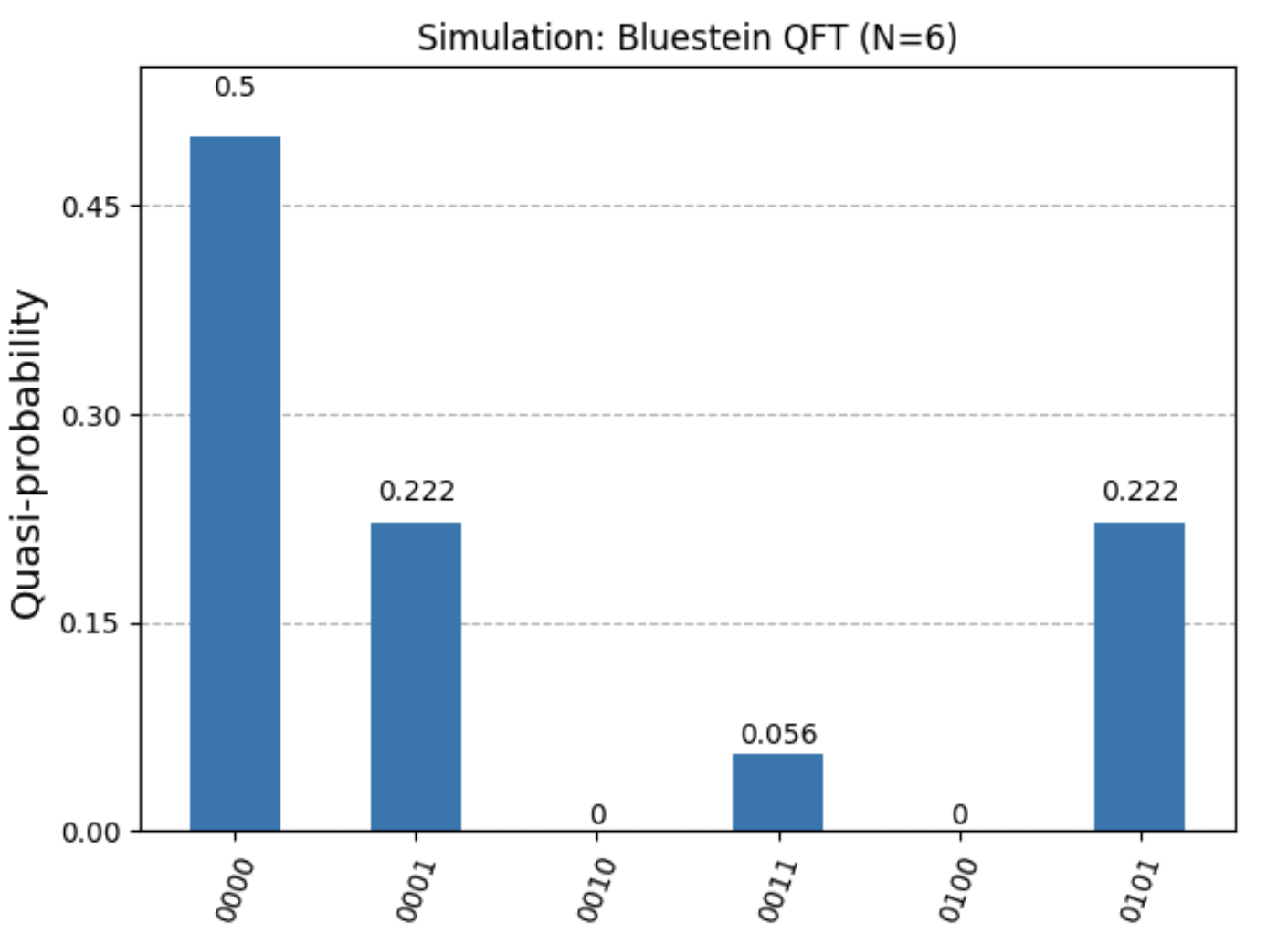}
\caption{Histogram of probabilities for $N=6$ generated using QBA.}
\label{fig-qba-n6-hist}
\end{figure}

Within the valid logical subspace (the first $N=6$ computational basis states), the output matches the target QFT state exactly. 
Fig.~\ref{fig-qba-n6} shows the corresponding circuit, while Fig.~\ref{fig-qba-n6-hist} shows the histogram. 

\section{Acknowledgments}

Renata Wong acknowledges support from the National Science and Technology Council grant No. NSTC 114-2112-M-182-002-MY3 and from Chang Gung Memorial Hospital grant No. BMRPL94.



\end{document}